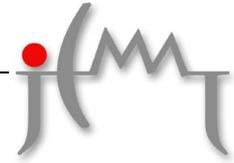

Research Article

# OSM: Leveraging model checking for observing dynamic behaviors in aspect-oriented applications

**Anas Mohammad Ramadan AlSobeh** [1,2*]

0000-0002-1506-7924

[1] Department of Information Systems, Faculty of Information Technology and Computer Science, Yarmouk University, Irbid, JORDAN
[2] Program of Information Technology, School of Computing, College of Engineering, Computing, Technology, and Mathematics, Southern Illinois University, Carbondale, IL, USA
* Corresponding author: anas.alsobeh@yu.edu.jo

## INTRODUCTION

Aspect-oriented programming (AOP) is an innovative paradigm in software development that aims to enhance the modularity of software systems by encapsulating cross-cutting concerns, which are functionalities that span multiple modules, into distinct entities called aspects that can then be woven into the core program. This can make the code more modular and easier to maintain (Abdulhameed, 2020; AlSobeh, 2014). However, AOP can also make it difficult to predict how a program will behave, and to test and verify its correctness. This is because the dynamic weaving of elements into the underlying code can result in complex interactions, posing challenges in predicting system behavior and maintaining accuracy (AlSobeh et al., 2020).

Despite AOP's effectiveness in achieving optimal modularity and separation of concerns, verifying the correctness of a woven program remains a nontrivial task. This is because creating a dynamic model that considers various factors such as aspect interactions, weaving processes, and system requirements, increases the complexity of the overall system (Alsobeh et al., 2020; Cheers & Lin, 2021). Therefore, handling large-scale systems with numerous aspects and complex interactions could result in increased computation time and resource requirements that affect the system's overall performance, particularly in time-critical applications.

To deal with such drawbacks, improving the dependability, safety, and flexibility of software systems, especially in critical sectors like electronic health records (EHRs), may be accomplished by creating a dynamic model and a related framework for testing and validating aspect-oriented (AO) models in weaved applications. AlSobeh et al. (2020) presented modular ontology model, which integrates AOP and ontology-based to develop dynamic EHR-aware services (Alsobeh, 2019).

Thus, this necessitates the development of a dynamic model that balances optimal results and woven program accuracy while devising a framework to verify and validate the behavior of AO models in the woven application, ultimately achieving dynamic quality model detection (Harel, 2020; Zhang, 2023).

Our approach leverages model checking to synthesize test cases for verification purposes. Model checking is a formal method that analyzes all possible states of a system to verify conformance to specified properties, enabling exhaustive analysis of correctness (Qu et al., 2021). By applying model checking to generate targeted test cases, we can effectively expose defects in the implementation under examination with respect to desired attributes. This formal, systematic approach is well-suited for assessing behavioral conformance in concurrent systems exhibiting finite state spaces. The capabilities of model checking facilitate verification of diverse behavioral properties through exploration of the entire state space (Karna et al., 2018). We utilize model checking to synthesize test cases likely to reveal defects in the system under examination, rendering a formal methodology apt for observing dynamic behaviors of AO systems. Specifically, model checking enables exhaustive state space analysis to automatically verify correctness properties of concurrent systems exhibiting finite state spaces. By modeling all possible state transitions, model checking can methodically generate targeted test scenarios aimed at exposing implementation issues relative to specifications. This systematic approach facilitates automated verification of temporal logic properties and assertions about ordering of events and states. Overall, model checking constitutes an effective formal technique for monitoring and validating the intricate dynamic behaviors and interactions prevalent in AOP systems (Xu et al., 2009).

To enhance quality metrics and streamline testing, AOP soundly modularizes cross-cutting concerns into cohesive aspects decoupled from core business logic. Specifically, AOP enables clean separation of orthogonal non-functional requirements into modular aspects that localize and encapsulate cross-cutting functionality. These aspects can be independently validated in isolation from the core program flow. Decoupling through aspects increases testability, measurably improving code quality metrics by localizing complex, tangled logic and minimizing scattering across disparate modules. Aspects form abstract, reusable modules that help disentangle cross-cutting concerns from primary business functions. This separation of concerns via quantification and obliviousness principles fundamentally facilitates testing and metrics analysis, reducing defects and technical debt. Overall, AOP's principled decomposition mechanisms profoundly augment modularity, promoting improved testing, measurement, and quality assurance (Gulia, 2019).

Therefore, the principled decomposition and modularization of cross-cutting concerns enabled by AOP quantifiably improves code quality metrics by localizing complexity and reducing coupling. Specifically, the separation of tangled logic into cohesive, independently testable aspects yields measurable reductions in defects and technical debt accrual. By disentangling and encapsulating cross-cutting functionality into modular units, AOP fundamentally curtails bug propagation and architectural erosion. The oblivious quantification of scattering and tangling via modular aspects has been empirically demonstrated to curtail defect density and improve reusability. Hence, the judicious application of AOP's core principles of separation of concerns, quantification, and obliviousness provides observable boosts to quality indicators through heightened modularity and reduced coupling. The ensuing improvements in cohesion, complexity management, and independent testability confer definitive benefits in terms of reduced bugs and defects (AlSobeh & Shatnawi, 2023; Cerone et al., 2021).

When combined with statistical model checking approaches, software models used during runtime allow automatic reasoning about system changes, identification of harmful or dangerous configurations, and possible self-adaptation (Idate, 2023). Due to the limitations of in-memory processing, traditional statistical model verification methods and tools may not be immediately relevant during execution. By fusing statistical model checking with AOP, we close the gap between weaving cross-cutting issues into a novel application, which is the main barrier of applying statistical model verification during program execution. Thus, it is critical to build an observe-based statistical model-checking (OSM) framework, to evaluate program behaviors during weaving or injection of new features (e.g., cross-cutting concerns) to get the object-oriented program's (OOP's) context data (Marquez et al., 2023).

Decoupling concerns that often overlap, such as patient data management, access control, and data security, improves an EHR system's modularity and maintainability (Hung, 2019). Model checking verifies the correctness of such components, hence enhancing the reliability of EHR system. To continue to serve their intended purpose, EHR software systems in dynamic environments need regular updates. Vaccination Management, vaccine inventory, patient scheduling, checking of adverse effects, and reporting vaccination status to public health organizations are just a few examples of areas, where inflexible EHR systems lead to undesired results. The robustness, security, and flexibility of EHR systems in dynamic settings ensured via the use of model checking to test the accuracy of interactions between these components and the main application logic. An AOP-based observer-based model checking framework is more robust than OOP while processing corrupted input, which may occur during program weaving or the introduction of additional cross-cutting concerns (Alsobeh et al., 2018).

AOP is used in EHR systems to modularize cross-cutting issues in a way that keeps them isolated from the primary application functionality. Adding an element requires encapsulating the fundamental features of an app so that you may change or add new ones. The interplay between the central functions and the injected features are affected by the modifications. To guarantee the dependability and consistency of the resultant EHR system, it is essential to understand these features and their influence on the weaving process. The proposed study is focused on the development of a runtime-observing system that is suitable for AO systems and is supported by statistical model verification techniques. This strategy employs formal approaches to ensure the reliability and precision of the observing procedure, hence boosting the performance and safety of AO systems in real-world circumstances. Moreover, the use of software models and statistical model checking techniques at runtime may facilitate automated reasoning regarding system changes, detect harmful or risky configurations, and potentially enable appropriate self-reactions. However, traditional statistical model checking techniques and tools may not be directly applied at runtime due to constraints related to execution time and memory occupation imposed by on-the-fly analysis, which is a certain issue may also emerge when utilizing AOP (Aichernig et al., 2019; Alsobeh et al., 2018; Nordine, 2023).We propose an innovative OSM framework leveraging AOP to enhance reliability and adaptability of EHR systems in dynamic environments. OSM framework uniquely integrates systematic pipelines for parsing, formal verification, computational analysis, and validation. This enables rigorous monitoring of program behaviors during weaving or injection of new features (Qader, 2022).

A key innovation is the use of a propositional model to observe EHR system properties and constraints. This model facilitates formal reasoning about safety, robustness, and adaptability amid continuously evolving requirements. OSM framework constitutes a novel amalgamation of statistical model checking, AOP modularity, and formal methods for verifying EHR system correctness. It provides a much-needed way to rigorously validate reliability and seamlessly adapt these mission-critical software systems to new conditions and regulations. The unique synthesis of modular AOP implementation, automated formal verification, and propositional modeling of constraints represents an innovative contribution. This novel approach drives the field forward in managing complex, evolving EHR software systems in a methodical, validated way.

## LITERATURE REVIEW

Cross-cutting concerns are like a ball of yarn that has become twisted as systems have become more complex. This metaphor can be used to understand the difficulties of addressing cross-cutting concerns in complex systems (Abid et al., 2022; Besser, 2019). Just as it is difficult to unravel a knotted ball of yarn, it is also challenging to apply cross-cutting concerns in an efficient and effective manner. This is because cross-cutting concerns have effects on a variety of different parts of a system. Several approaches, including code duplication, error handling by hand, and manual security checks, have been used in the past to address cross-cutting problems. But nonetheless, these methods are notorious for their complexity and frequent breakdowns (Alsobeh et al., 2019). OOP is not well-suited for implementing cross-cutting concerns. This is because cross-cutting concerns often affect multiple objects in a system. OOP does not have a good way of dealing with this. AOP has been appeared as a promising approach to provide better separation and integration of cross-cutting concerns than plain OOP (Nusayr et al., 2022). AOP can also help to improve the quality, maintainability, and testability of OOP code. It is considered as a complement for OOP, not as a replacement to it, which provides a way to implement cross-cutting concerns in a consistent and reusable way (Alsobeh & Clyde, 2014).

### Aspect-Oriented Programming & Model Checking Software Applications

AOP is a programming paradigm that encapsulates cross-cutting concerns into aspects with the appearance of specified points in program execution. It has many benefits such as enhancement of reusing and changing to create more value for software system developers and users. There are two sides to using aspects: the good side is the enhancement of reusing, and the risk side it can be used in a harmful way that can break the integrity concept of the programs (Abdulhameed et al., 2020; Ghareb & Allen, 2018; Patel et al., 2023). During our work, we focused on investigating, where developers can use AOP in the software development process focusing on cross-cutting concerns especially on two important phenomena: tangling and scattering. A simple advice can change the whole behavior of the base classes whatever it is expected or unexpected. Aspect must be used and applied in an accurate and effective way. A strict analysis must be used to ensure the quality of AO system. It supports separation of cross-cutting concerns by building a new unit of modularization, which called an "aspect". Every aspect has its own cross-cutting functionality (AlSobeh & Clyde, 2014; Georg et al., 2009). This will decrease the heavy load on the core classes. This will decrease the heavy load on the core classes. An aspect weaver creates the final system by joining and gathering the core classes within cross-cutting aspects through a process called "weaving". This weaving together cross-cutting concerns into a single, cohesive system, which allows developers to decouple cross-cutting concerns from the main business logic. This makes it easier to test and maintain the software.

New features, including logging, safety, and efficiency, were offered by Xu et al. (2007). It is possible for issues to arise if any of the system's components are not compatible with the overall design. They do not add unnecessary complexity, slowness, or vulnerability to the system. They showed how model checking may be used to ensure that the attributes of your system are not altered by any of its components. Model checking explores every potential situation while analyzing a system's behavior. Before committing to a system's full implementation, this might help you identify potential issues. They demonstrated how to adapt these models for use with LTSAs. LTSA can guarantee that the models work with your infrastructure. The study's findings suggested that this approach may be used to ensure the validity and high standard of future AO designs. By illuminating dependencies and making them more manageable, it may also help in problem-solving. Dynamic weaving and parameterized pointcuts, two notable exceptions, are not available due to the need for manual model conversion into LTSA processes. It does not judge superficial factors like processing speed or RAM use.

Xu et al. (2022) introduced a method for modeling and verifying AO systems using finite state machines by encapsulating aspects and their interactions with classes into FSP processes and testing its efficacy by identifying faulty models. Some sophisticated capabilities of AO languages, including dynamic weaving and parameterized pointcuts, are not supported in the paper. Another shortcoming is that it does not consider the effect of factors on operational attributes like speed or memory use. Thirdly, turning models into FSP processes is labor-intensive and hence limited.

Alsobeh and Clyde (2016) demonstrated the potential of AOP to implement transaction-related cross-cutting concerns in modular, cohesive and loosely coupled transaction-aware aspects, by proposing TransJ framework, which provides join points and pointcuts for weaving advice into high-level runtime abstractions, such as transaction contexts. Thus, cloud computing, e-commerce, online banking, social media, and big data analytic, dynamic analysis are all some of current challenges that can be tackled using AOP since they all involve dispersed transactions.

### Model Checking of Data Systems

A well-documented issue in model checking is the state space explosion problem, which is particularly exacerbated in the realm of big data, as it is characterized by a significant increase in the quantity, diversity, and speed of data (Cerone, 2021). The presence of ex-tensive data volumes results in an exponential growth in the potential states that a system can occupy. Consequently, this gives rise to a state space of considerable magnitude that necessitates exploration, which can be impractical or unfeasible. The issue is exacerbated by the existence of intricate data structures and types. Distributed architectures are frequently employed in big data systems, thereby introducing an additional level of intricacy to the process of model checking. The complex interplay among various components and data distribution mechanisms poses significant difficulties in constructing a comprehensive and precise model of the system. Consequently, the applicability of conventional exhaustive model checking methods, which demonstrate effectiveness in small-er and less intricate systems, may be limited in the context of big data systems. Nevertheless, the application of model checking to big data software poses various challenges. The phenomenon of state space explosion is a fundamental concern in the field, as the inclusion of all conceivable data values results in state spaces of unmanageable magnitude. The problem has been addressed through the study of abstraction and modular verification techniques. Holistic system modeling and verification encounter additional challenges when dealing with data-intensive distributed architectures. The optimization techniques of runtime monitoring and selective validation have been proposed in the literature (Camilli, 2014).

### Abstraction and Modular Verification

Abstraction involves reducing the level of detail in the system model to focus on relevant behaviors and properties (Grumberg, 1994). Data abstraction can be used to group concrete data values into abstract representations. Control abstraction simplifies complex control logic into atomic transitions. Environment abstraction models only the external behaviors of subsystem components. Appropriate abstractions allow faster verification by reducing the state space. However, abstraction must balance precision and analysis speed. Coarse abstractions may miss bugs while overly detailed ones negate computational benefits. Automated support for generating and refining abstractions is an active research area (Camilli et al., 2014).

### Modular Verification

Modular verification decomposes the analysis of large inter-connected systems into modules that can be verified independently (Grumberg, 1994): compositional verification of parallel programs (CVPPs) and formal methods in computer-aided design. This exploits the locality of interactions to alleviate state space growth. Individual modules are model checked, then an incremental composition process formally derives system-level properties. Lightweight synchronization models and interface contracts between modules enables scalable verification. Challenges include handling complex interfaces and discrepancies between local and global specifications. Integration with development workflows is also required for adoption. Projects like CVPPs are advancing modular techniques for real-world systems. By care-fully applying abstraction and modularization, model checking can be scaled to verify critical behavioral properties of big data systems. The ability to provide sound guarantees of correctness makes this a promising area for continued research.

Cheers et al. (2021) presented AO state machines are a type of state machine that can be used to model the dynamic behavior of AO programs. The extension to UML provides a number of new features that can be used to model the dynamic behavior of AO programs, including the ability to model the injection of new code into existing classes, the ability to model the interaction between aspects and classes, and the ability to model the dynamic behavior of aspects. However, they did not verify the correctness of the models that are created using the extension to UML. Other research are limited in its scope and does not provide any formal methods or empirical evidence to support the claim that the extension to object-oriented is effective in modeling the dynamics of AO programs.

## OBSERVE-BASED STATISTICAL MODEL-CHECKING ARCHITECTURE

To construct the proposed OSM framework using the quality detection model depicted in **Figure 1**. It shows OSM architecture to validate the behavior of complex systems based on observed traces or executions. Rather than constructing a mathematical model, this approach leverages real-world data for verification purposes. OSM framework champions a paradigm, where 'observing and learning' from real-world system operations can lead to more accurate, robust, and adaptive system validation, especially when dealing with complex and dynamic systems. **Figure 1** offers a graphical representation of OSM framework. This is not just an abstract diagram but a roadmap to how the system operates. Here's a step-by-step walk-through:

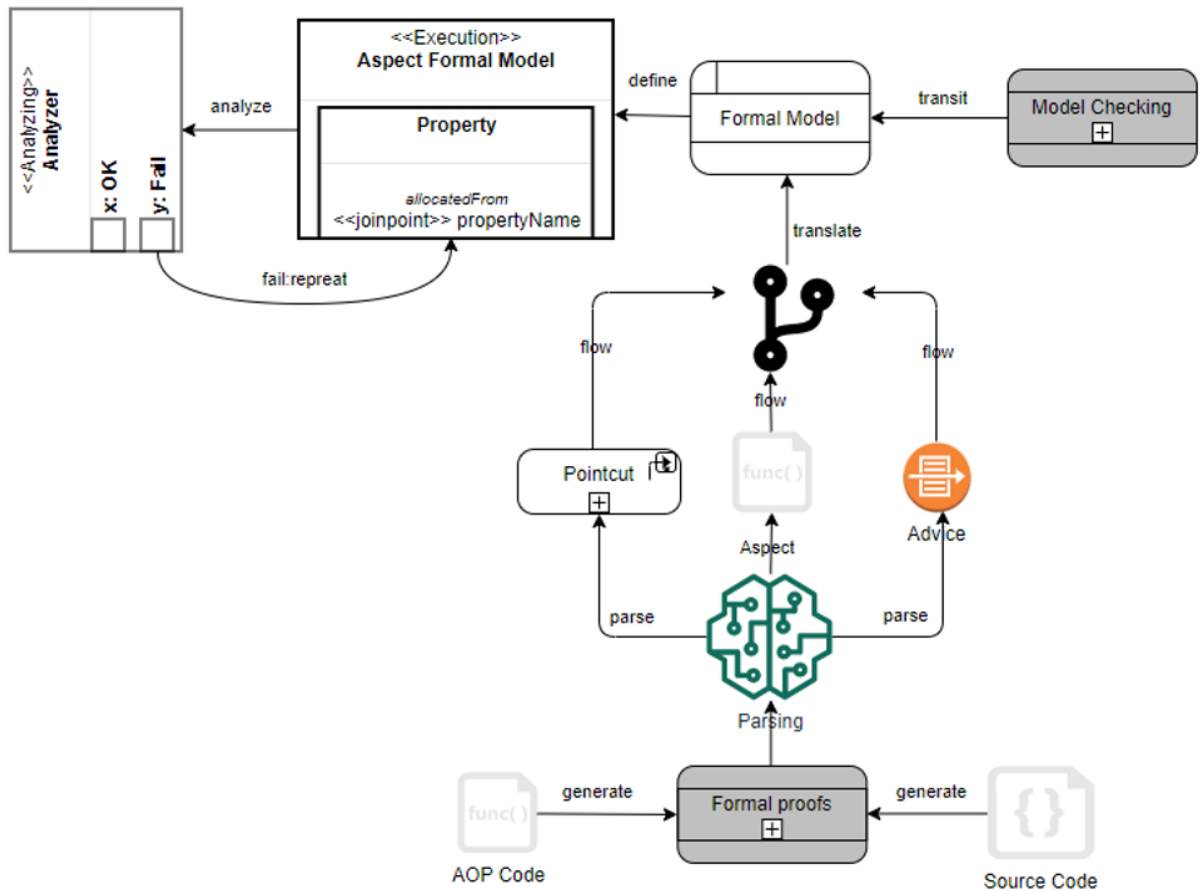

**Figure 1.** OSM architecture & workflow (Source: Author)

1. **Data collection:** As a system operates, it constantly generates data–logs of actions taken, outputs produced, errors encountered, and more. This stream of information is the raw material for OSM. In the context of a software application, these might be logs generated by servers, user interactions recorded in databases, or system performance metrics.
2. **Behavioral tracing:** Within the collected data, specific patterns of behavior emerge. These patterns, or 'traces,' are sequences of events that recur, reflecting how different components of the system interact under varying conditions.
3. **Framework construction:** OSM framework ingests these traces, organizing and classifying them. It's akin to piecing together a puzzle, where each trace offers insight into a different part of the system's overall behavior.
4. **Validation & verification:** Armed with this organized dataset, OSM can then compare the observed behavior (from the traces) against expected behavior (perhaps defined by system specifications or benchmarks). Any discrepancies here indicate areas that might need further investigation or refinement.

In a simple scenario, imagine a smart city traffic management system. Every day, sensors at intersections record millions of data points: car counts, traffic light changes, pedestrian crossings, and more (Tashtoush et al., 2022). Traditional modeling might struggle to predict traffic flow during a surprise event, like a parade or a power outage. However, OSM framework, by continuously observing and learning from the city's actual traffic data, can recognize unusual patterns and adapt more quickly. If a similar event has happened before, even if it was a minor one, OSM would have recorded the trace and can predict potential outcomes, offering traffic solutions in real-time.

OSM generates formal models automatically from the source code. This entails parsing code to determine the essential elements, such as aspects, pointcuts, and advice. This is possible with an AO parser that comprehends AOP's special syntax and structures. After identifying the components, construct a CFG for each module, considering both the base code and the aspects. CFG represents the control flow between the various components, including any possible interactions between aspects and base code. After constructing CFGs, translate them into a formal model suitable for model verification. This includes translating CFGs into labeled transition systems, such as Kripke structures (Hachani et al., 2002), or any other formalism that model checking tools can use to verify the properties of the system. After formal models have been generated, aspect formal checking can be per-formed, which specifies the properties that the system is expected to meet, such as safety, liveness, and fairness properties. These properties can be expressed in temporal logic or any other supported formalism by executing the model checking aspect process, which entails analyzing the generated formal models and determining if the specified system properties hold, such as expressing desired properties in temporal logic, e.g., linear temporal logic (LTL) or computation tree logic (CTL) (Zhu, 2021). If the properties do not hold, the model checker will provide an alternative on the formal models and the temporal logic properties (Xu, 2007). These properties can be described as propositional variables and construct logical expressions to describe the relationships between them. Analyzing the results to identify potential issues or confirm that the system satisfies the desired properties, which means the output of the analyzer may discover any problems in AOP code or the interactions between aspects and the core code. If any flaws are found, the code must be modified, and the process must be repeated. This iterative cycle allows for continuous system improvement and validation, thus reinforcing the dependability of safety-critical and real-time systems.

### OSM Process: Enhancing EHR Systems Including COVID-19 Requirements

Today's healthcare landscape demands EHR systems that not only store patient data but also provide seamless access to numerous stakeholders, including doctors, nurses, researchers, insurance agencies, and medical students. EHR systems stand at the nexus of patient care, ensuring safe and compliant data management.

The intricate design of an EHR system developed through AOP involves foundational functionalities (P) for patient data management and encompasses cross-cutting concerns. These include access control (A), data privacy (B), health service support (C), vaccination management (D), encryption (E), and logging (L).

To illustrate the intricate process of transforming source code into a formal model, consider the functionality for patient data management (P). The system's source code dictates how data is stored, retrieved, and modified. When transforming this into a formal model, each function or method in the source code is mapped to a specific state or transition in the model. For instance, adding new patient data might translate to a transition from a state "no data" to "data available". Such transformations are executed for all functionalities, ensuring that the formal model accurately represents the entire system's behavior.

In the age of pandemics, it is paramount for EHR systems to integrate new medical findings, diagnostics, treatments, and patient outcomes without disruptions. By allowing dynamic modification of data gathering, reporting, and analytical procedures, the system can closely monitor the evolving trajectory of diseases like COVID-19.

The overarching goal of a state-of-the-art EHR is ensuring safe, secure, and real-time sharing of patient information among healthcare entities without undermining patient privacy. Especially during pandemics, this facilitates a unified data approach, fostering collaboration and enhancing response efficiency.

Key features of a modern EHR include provisions for remote patient consultations, streamlined medication management, and encrypted channels for communication between healthcare providers and patients. Amplifying its health service support functions can further augment the system's ability to cater to surging demands for remote medical care.

The intricacies of COVID-19 vaccination distribution, tracking, and administration are efficiently addressed by an adept EHR system. As detailed by Khalifa (2020), the system's vaccine management feature is enhanced to oversee vaccine inventory, patient appointments, monitor side-effects, and relay data to health agencies (Khalifa, 2020). Moreover, throughout a crisis like the COVID-19 pandemic, adherence to access control and privacy protocols (A, B) remains critical, requiring systems to be adaptable to emerging guidelines or legislative mandates. Ensuring enhanced access control (A) and data privacy (B) mechanisms, EHR system becomes a pivotal tool in managing patient information during the pandemic, aiding healthcare professionals in delivering top-notch care and effectively responding to multifaceted challenges.

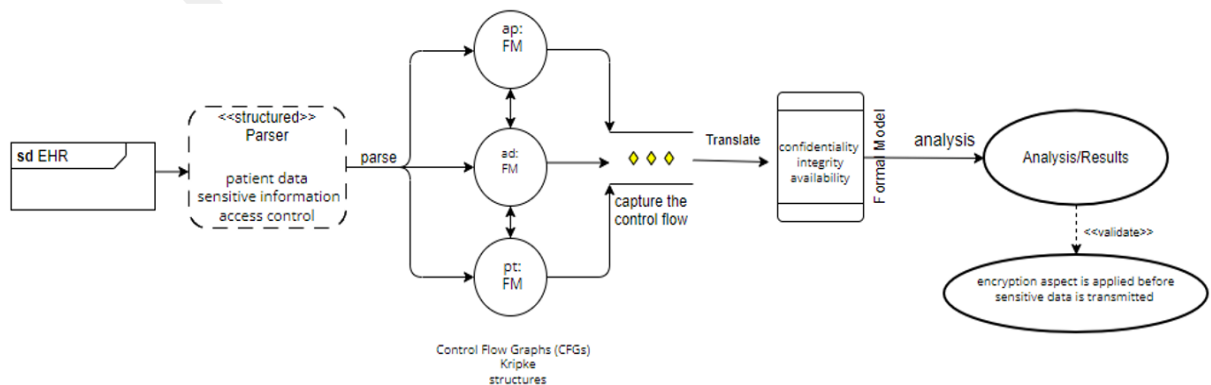

**Figure 2.** Interaction of an EHR system within OSM processes (Source: Author)

The various stages of integrating elements like logging, encryption, access control, and others into the core functionality of EHR system through the lens of big data systems can present some complexities. **Figure 2** lies in its clear demonstration of the systematic integration process, highlighting the comprehensive steps involved in OSM's processes and underscoring how they enhance the robustness and quality of EHR system. It illustrates the applying OSM process to EHR system, which helps ensure that the desired properties are maintained, enhancing the safety, reliability, and overall quality of the system. Suppose we have an EHR system developed using AOP that includes aspects for logging patient data, encryption of sensitive information, and access control. In EHR system with the following modules:

1. Patient data management (P)–the core functionality.
2. Access control (A), data privacy (B), health service support (C), vaccine management (D), encryption (E), and logging (L)–the cross-cutting concerns.

Each of these modules are described as propositional variables. OSM weaving process uses logical expressions that combine these variables, where the complete EHR system (S) can be represented as the conjunction of the core functionality and the cross-cutting concern expression (proposition 1):

$S = P \wedge A \wedge B \wedge C \wedge L \wedge E$.

Proposition 1 defines predicate on the relationships between these components using propositional logic. For instance, P(x) represents patient data management for patient x, while A(x) signifies access control for patient x, and so on. The complete EHR system for a patient x, incorporating all the cross-cutting concerns as predicates, which can express interactions and constraints between various components of EHR system. Then we can verify several universal interactions, i.e., proposition 2. The complete EHR system for a patient x, incorporating all the cross-cutting concerns as predicates, as shown in proposition 1, which can express interactions and constraints between various components of EHR system. Then we can verify several universal interactions, i.e., proposition 2:

$\forall x \ (C(x) \rightarrow A(x) \wedge B(x))$.

If health service support (C) is present for a patient x, then access control (A) and data privacy (B) must also be present for that patient x (proposition 3):

$\forall x (A(x) \rightarrow (L(x) \wedge E(x)))$.

If access control (A) is present for patient x, then logging (L) and encryption (E) must also be present for that patient x. These expressions allow us to describe the system's properties and constraints as model checking. For example, if health service support (C) requires specific access control (A) and data privacy (B) functionalities, we can stand for this constraint, as follows:

$C \rightarrow (A \wedge B)$.

This expression states that if health service support (C) is present in the system, then access control (A) and data privacy (B) must also be pre-sent. Another example, if access control (A) requires specific logging (L) and encryption (E) functionalities, we can show this predicate, as follows:

$A \rightarrow (L \wedge E)$.

```
package OSMAspect;
public aspect LoggingAspect {
// Pointcut: capture all methods related to patient data management
pointcut patientDataOperations(): call(* com.example.ehr.PatientData.*(..));
// Advice: log the start and end of each patient data operation
before(): patientDataOperations() {
System.out.println("Starting operation: " + thisJoinPoint.getSignature());
}
after(): patientDataOperations() {
System.out.println("Finished operation: " + thisJoinPoint.getSignature());
}
}
public aspect EncryptionAspect {
// Pointcut: capture all methods that handle sensitive patient information
pointcut sensitiveDataOperations(): call(*
com.example.ehr.PatientData.get*(..));
// Advice: encrypt sensitive data before returning it
Object around(): sensitiveDataOperations() {
Object result = proceed();
return encrypt(result);
}
private Object encrypt(Object data) {..}
}
public aspect AccessControlAspect {
// Pointcut: capture all methods related to patient data access
pointcut patientDataAccess(): call(* com.example.ehr.PatientData.*(..));
// Advice: check user's authorization before allowing access
before(): patientDataAccess() {
if (!isUserAuthorized()) {..}
}
private boolean isUserAuthorized() {..}
}
package EHR;
class PatientData {
public MedicalHistory getMedicalHistory(int patientID) {
// Fetch patient's medical history from the database}
}
```

**Figure 3.** Snippet code of key aspects & core functionality in EHR system (Source: Author)

This OSM's logic states that if access control (A) is present in the system, then logging (L) and encryption (E) must also be present. Similarly, we can describe other cross-cutting concerns and interactions between EHR components in OSM using propositional logic and formal methods. The intersection between these modules using basic mathematical operations for injection, where translate, if C, then A and B must also be present for the patient, $C(x)=A(x)+B(x)$, then L and E must also be present for that patient, $A(x)=L(x)+E(x)$. OSM's analysis processes logical expressions to verify properties of these interactions, such as whether certain components can coexist or whether specific cross-cutting concerns are adequately addressed.

**Figure 3** shows the snippet code of EHR cross-cutting concerns, including the core functionality (P) and the aspects for L, E, and A, is parsed. The snippets offer a clear and practical illustration of how these critical aspects are implemented in EHR core code using OSM's components, where it contributes to the cross-cutting functionalities of EHR system, such as the Logging Aspect helps track system activities, the Encryption Aspect ensures the security of sensitive patient data, and the access control aspect controls who has the authorization to access specific resources. Core functionality, represented by PatientData.getMedicalHistory (P), demonstrates how patient medical history is retrieved in the system. OSM demonstrates the practical application and interplay of these aspects within the core code of an EHR system, helping readers understand the real-world implementation of such crosscutting concerns. The parser identifies the aspects, pointcuts, and advice in the core code, creating a representation of AOP structure of EHR system, such that implements three aspects: L, E, and A, as shown in **Figure 3**.

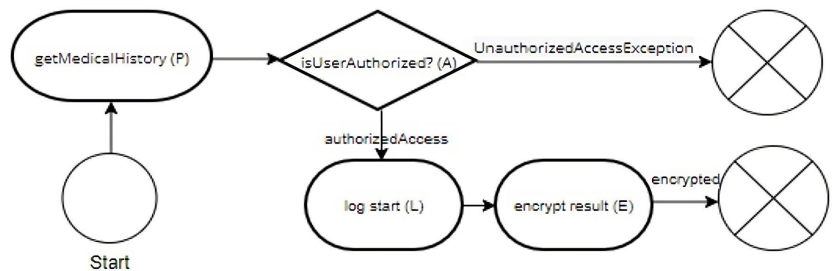

**Figure 4.** CFGs, interplay of core functionality & aspects (L, E, & A) in patient data management (Source: Author)

These aspects are woven into EHR system, enhancing the core functionality related to P. The parser reads the source code, extracting the aspects, pointcuts, and advice to create a representation of AOP structure of EHR system. This representation is then used to construct CFGs and, subsequently, formal models for model checking.

OSM framework automates the generation of formal models directly from system source code, enabling rigorous verification. This involves systematically parsing the codebase to extract key structural elements, including aspects, join points, pointcuts, and advice in AOP implementations. For example, an AspectJ parser would analyze the abstract syntax tree (AST) to identify aspect declarations, before/after advice, etc.

With AOP structure identified, the next stage constructs CFGs to represent execution paths within and between components. Nodes in CFG denote statements or expressions, while edges capture flow of control. CFG visualizes possible flows considering aspect interactions with the base code.

For example, a CFG for an EHR getMedicalHistory call could show flows from checking user access permissions, to logging entry/exit, to fetching encrypted records. Loops, branches, and entry/exit points are modeled. CFG for getMedicalHistory would translate to a Kripke model representing those components and their interactions, like state 1: start, transition 1→2: isUserAuthorized (A), state 2: log start (L), transition 2→3: encryptData (E), state 3: fetch records (P), transition 3→4: log end (L), and state 4: return records.

This formal representation enables exhaustive verification of EHR system correctness and security properties through model checking.

CFGs should capture the control flow between the aspects and the base code, reflecting the proper execution order and any interactions. In considering the PatientData class (core functionality) and the three aspects (L, E, and A) from the earlier response, as shown in **Figure 4**. It captures the interactions and dependencies between them. In this CFG, the nodes represent the various operations and method calls, while the edges represent the flow of control between these operations. The graph begins with the getMedicalHistory method call (core functionality P) and proceeds to the isUserAuthorized check (access control aspect A). If the user is unauthorized, an UnauthorizedAccessException is thrown, and the control flow ends. If the user is authorized, the control flow proceeds to the log start operation (logging aspect L), followed by the encrypt result operation (encryption aspect E). Finally, the control flow moves to the log end operation (logging aspect L) and then returns the encrypted result. The graph can then be translated into a formal model suitable for model checking, such as labeled transition systems or Kripke structures.

The formal models are checked against the desired proper-ties and constraints defined using propositional logic (e.g., A→(L∧E)). It defines proper-ties EHR system should satisfy, such as confidentiality (e.g., sensitive data must al-ways be encrypted), integrity (e.g., datashould not be modified without proper authorization), and availability (e.g., authorized users should always have access to the system). Run the model checking process on the generated formal models, verifying that the specified properties hold. For example, the model checker may verify that the encryption aspect is always applied before sensitive data is transmitted or stored. If the model checker con-firms that the properties hold, EHR system is considered safe and reliable. However, if any issues are detected (e.g., a counterexample showing unauthorized access to sensitive data), EHR system's AO code should be revised to address the problem. The model checking process is then repeated to ensure that the desired properties hold after the modifications. Therefore, the results of the analyzer verify if EHR system adheres to the specified constraints and requirements, ensuring the correct implementation and interaction of the aspects within EHR system.

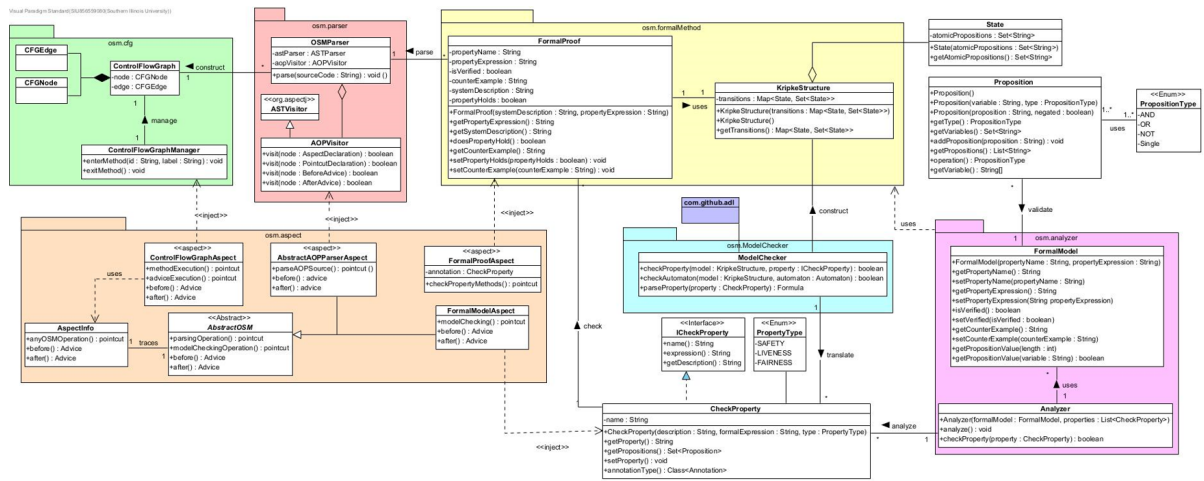

**Figure 5.** OSM framework components (Source: Author)

This shows AO captures the complexity and relationships between the modules in the actual EHR system using directed graph representation. In our EHR System: we directed edges between the nodes to represent dependencies and relationships: C→A: If health service support (C) is present for a patient x, then access control (A) must also be present for that patient x. C→B: If health service support (C) is present for a patient x, then B must also be present for that patient x. A→L: If A is pre-sent for patient x, then L must also be present for that patient x. A→E: If A is present for patient x, then E must also be present for that patient x. However, it's possible that this model is insufficient on its own for ensuring the system's integrity and security. Formal techniques, model checking, or testing may be used to conduct a more thorough analysis of the system. These methods may boost confidence that system will provide the expected functionality, safety, and dependability.

## OSM'S COMPONENTS & IMPLEMENTATION[1]

**Figure 5** shows the integrated pipeline of parsing, model checking, formal computation, analysis, and validation, which enables the verification of system's properties, ensuring that components can coexist and that specific cross-cutting concerns are addressed appropriately. By analyzing logical expressions and executing the model checking process, EHR system's adherence to specified constraints and requirements is confirmed, guaranteeing the correct implementation and interaction of the aspects within the system.

OSMParser uses the visitor pattern to traverse and analyze AST representing AOP code. The Visitor pattern is a behavioral design pattern that allows you to define new operations on an object structure without changing the structure itself (Hachani et al., 2002; Pereira-Vásquez et al., 2020). It is particularly useful when dealing with a complex hierarchy of objects or when you need to perform multiple operations on the same object structure. AspectInfo class holds some basic information about an aspect, where the Aspectprocessor has a processAspects method that is retrieving the collected AspectInfo instances by creating an instance of AOPVisitor, parsing the AspectJ source code. OSMParser class is responsible for parsing the source code and generating an AST using ASTParser. AOPVisitor class is a visitor that extended by ASTVisitor class, which is the abstract visitor class provided by the Java AST framework. It can visit different nodes in AST and identify the essential AOP elements (aspects, pointcuts, and advice), which defines methods for visiting these nodes, such as visit (AspectDeclaration), visit (PointcutDeclaration), visit (BeforeAdvice), and visit (AfterAdvice). When a specific node is encountered during the traversal, the corresponding method in AOPVisitor class is called, allowing the visitor to perform the necessary analysis on that node. The AbstractAOPParserAspect is an abstract aspect that contains a pointcut called parseAOPSource () and two advice methods: before() and after(). It handles advising that is parsing and processing of AOP source code.

---

[1] https://github.com/siualsobeh/osm

```
Algorithm 1 OSM Model Check EHR System
 1: procedure MODELCHECKEHRSYSTEM(sourceCode)
 2:     aopStructure = ParseAOPStructure(sourceCode)
 3:     cfgs = ConstructCFGs(aopStructure)
 4:     formalModels = TranslateToFormalModels(cfgs)
 5:     properties = DefinePropertiesAndConstraints()
 6:     for all model in formalModels do
 7:         result = ModelCheck(model, properties)
 8:         if not result then
 9:             ReviseAOCode()
10:             Return ModelCheckEHRSystem(sourceCode)
11:         end if
12:     end for
13: end procedure
```

**Figure 6.** Algorithm 1: Procedure of OSM-DB framework for validating & ensuring safety & reliability of big data EHR software systems (Source: Author)

By using this aspect, you can modify the behavior of OSMParser and AOPVisitor without changing their actual implementation, i.e., obliviously, which reflects the modularity and extensibility benefits of AOP approach. Observer (visitor) pattern is used to traverse and analyze AST generated from AOP source code, while AOPVisitor class acts as the concrete visitor to identify essential AOP elements. The AbstractAOPParserAspect demonstrates the modularity of AOP approach by advising the parsing and processing of AOP source code. The ControlFlowGraphAspect is an aspect that handles weaving CFG generation into EHR target system. It intercepts the execution of the program at specific join points, such as method calls, branches, and loops, and updates CFG accordingly. This helps to modularize CFG generation process and separate EHR's flows from the core functionality of the system. The ControlFlowGraphManager class handles CFG construction for different modules and aspects in a program. It interacts with the parser to extract relevant information from the source code and AOP structures, and then builds the corresponding CFGs. The manager provides methods for accessing and manipulating the graphs, such as adding and removing nodes and edges, and querying the graphs for specific information. Thus, OSM's CFG classes enable flexible data-driven modeling of program control flow. The graph structure soundly captures statement relationships and execution trajectories within and across aspects. Systematic encapsulation and abstraction mechanisms provide versatility along with strong typing guarantees.

FormalProof class is responsible for conducting the formal verification process using CFG and the specified properties. It takes the generated CFG and translates it into a suitable formal model, such as a Kripke structure, which can be analyzed by model checking tools . The class also handles the expression of properties in temporal logic or any other supported formalism, such as LTL or CTL. CheckProperty class represents a specific property that the system is expected to be safety, liveness, or fairness properties, and are expressed in a formal language, such as temporal logic. This class also stores data about a property, such as its formal expression, type, and methods for comparing and evaluating the properties against the formal model derived from CFG, which can be checked to verify if the model satisfies the properties, and to identify any potential problems with the model. FormalProofAspect is an aspect's class that enables formal verification to be woven into the target program, which intercepts the execution of the construction of CFG or the translation of CFG into a formal model. It then invokes FormalProof class to perform the verification, checking if the specified properties hold for the system. Therefore, FormalProofAspect helps to modularize the formal verification process and separate it from the core functionality of the program. Thus, the analyzer ensures the correct implementation and interactions of the aspects within the system, and validates the logical relationships and constraints defined by the formal methods.

Algorithm 1 in **Figure 6** provides a method tailored for the construction and optimization of AOP-based EHR systems. Initiating the process, the raw AOP implementation of EHR system is parsed to generate an AST, illuminating the hierarchical structure and identifying key AOP constructs such as aspects, join points, pointcuts, and advice. This AST subsequently undergoes a transformation to produce CFGs, offering a granular representation of potential execution trajectories within the system, with nodes delineating AOP interactions and edges charting the execution sequence.

Following this, a Kripke transition system is derived from CFG, forming the foundation for rigorous model checking. Every node and edge from CFG translates seamlessly into states and transitions of the Kripke model, with added predicates to articulate the conditions governing each state. This methodical construction does not just enhance clarity but ensures conformance to formal specifications. Delving into the operational facet, the algorithm's efficiency is evident in its polynomial runtime complexity, predominantly steered by the intricacies of state space traversal in model checking. Complementing the core procedures, the algorithm boasts tight integration capabilities with prevailing AOP toolchains, a boon for developers, fostering modular design while preserving the flexibility to weave in new aspects or modify existing ones.

## EVALUATION & DISCUSSION

To evaluate our model, we used online datasets that contain EHR data from different sources and settings. One of the datasets we used was medical information mart for intensive care (MIMIC-IV), which was especially suitable for our model (Johnson, 2023). MIMIC-IV is a freely available dataset developed by the MIT lab for computational physiology. It includes de-identified health data of about 200,000 patients who stayed in critical care units between 2008 and 2019 (Johnson et al., 2023; Kallfelz et al., 2021). To access this dataset, we completed a CITI course in human research protections as a requirement to protect the privacy of the individuals whose data is in the database. We used it as input to simulate and test our model with healthcare systems, such as the patients seeking medical care via health service. We utilized this MIMIC-IV dataset to create a realistic and representative MIMIC-EHR system's extensions for testing the functionality of OSM model. They have attributes such as name, age, gender, address, phone number, email, medical history, etc. They have methods such as register, login, logout, request appointment, cancel appointment, join consultation, rate consultation, etc. This helped us evaluate the model's robustness and observe its behavior under realistic conditions and identify any potential issues. To facilitate a thorough and efficient healthcare response to the COVID-19 pandemic, we identified key cross-cutting problems that would support the evolving role of EHR systems. Although these are generic features included in many EHR systems, they have been tailored to meet the requirements of the COVID-19 EHR. When implemented, these functionalities are usually in the form of a module. These modules all fundamentally perform the COVID-19 functions of managing MIMIC-EHR system. However, these modules are commonly modified for COVID-19 EHR's domain-specific needs. For instance: Patient-centric care plans are automatically updated to reflect new information on how to treat or prevent COVID-19, thanks to this overarching concern. Data-driven, which is intersecting concern weaves data for real-time data interchange on COVID-19 vaccine development and dissemination by integrating with OSM-EHR components. In order to facilitate the delivery of healthcare remotely, telehealth is a cross-cutting problem that is integrated with OSM-EHR via the provision of seamless connectivity with telehealth platforms. Vaccine Management is a cross-cutting issue that keeps track of patients' COVID-19 immunization history, down to the vaccine kind and administration dates, and also includes a reminder and appointment setting function. These cross-cutting concerns can be integrated obliviously without changing the core EHR health system, as shown in **Figure 7**.

**Figure 7** demonstrates the integration of vaccine management, a critical cross-cutting concern that meticulously tracks patient COVID-19 vaccination histories, including details such as vaccine type and administration dates, in addition to a reminder and appointment setting functionality. Such cross-cutting concerns are incorporated obliviously into the core EHR health system using OSM's properties. Represented as individual modules in the MIMIC-EHR system, these cross-cutting issues demonstrate high extensibility and adaptability, vital for their integration into various health-focused application setups. Using a predefined EHR methodology to extract a set of EHR data, these modules are integrated into OSM framework, as showcased in this case study. These interdisciplinary issues are represented by individual modules inside a MIMIC-EHR system and must be highly extensible and adaptable to fit into any health-focused application setup. To extract a set of EHR data according to a pre-defined EHR methodology, we integrated these modules into OSM framework and utilized them in this case study, as shown in **Figure 7**. To evaluate these modules, we extracted relevant patient data from the MIMIC-IV dataset and simulated various tasks within OSM health service health system (EHR_MIMIC_Sys). This enabled us to assess how the system performed under realistic conditions and whether it met all the defined properties and constraints. The data we extracted included:

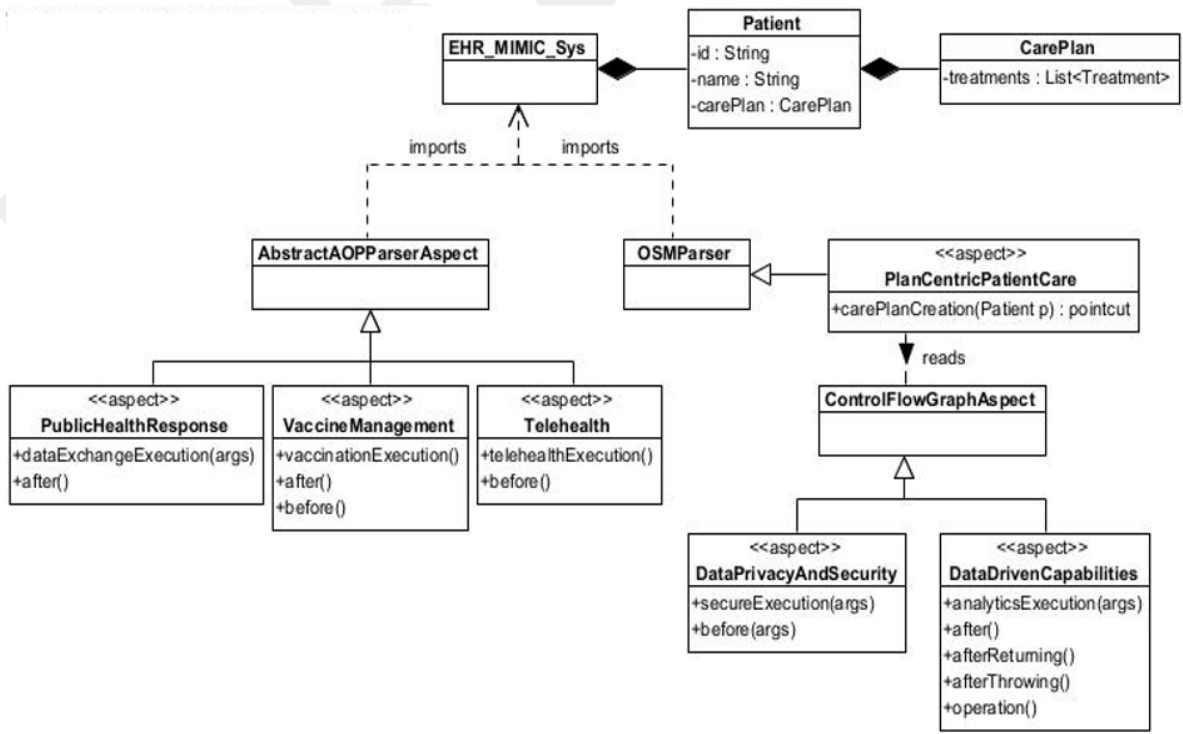

**Figure 7.** Integration of cross-cutting concerns into EHR system using OSM's aspects (Source: Author)

1. **Patient demographics:** Information such as age, gender, ethnicity, etc.
2. **Clinical data:** Lab measurements, medications, vital signs, diagnoses, etc.
3. **ICU data:** Caregiver notes, imaging reports, fluid balance, severity scores, etc.
4. **Hospital data:** In-hospital mortality data, hospital, and ICU length of stay, etc.

In our model, we generated AOP code for healthcare systems using the MIMIC-IV dataset as a basis. We then fed this AOP code into the model for parsing and analysis, as shown in **Figure 7**. We extended EHR system with OSM aspects that ensure weaving the system and its data obliviously. It uses pointcuts to intercept the patient records, such as applies encryption and decryption algorithms. It also uses pointcuts to intercept the create, retrieve and store methods of consultations. This required introducing the aspects to EHR system's classes, which created an advises relationship with the HealthResponse, VaccineManagement, Telehealth, DataPrivacyAndSecurity, DataDrivenCapablities, and PlanCentericPatientCare aspects. These aspects advise OSM framework's classes within the execution of the program. They identify a pointcut for the call and extend of the modelChecking, checkPropertyMethod, ParsingOperation, methodExecution methods and declare advice acting around that pointcut. They also ensure the personal and medical information of patients and providers is supplied with appropriate behavior, as shown in **Figure 7**. These aspects firstly extend a pointcut targeting the call of OSMParser, AbstractAOPParserAspect, andControlFlowGraphAspect. Secondly, the aspects implement advice that wrap the declared pointcuts to intercept and sort the returned flow and data. To document this application, its behavior and structure are modeled for requirements traceability using AspectInfo, while also applying the proposed FormalModelAspect. These aspects have also introduced new EHR System settings that stores a list of FormalProof class that need to be validated. In a complex EHR application, this could be represented as EHR configuration components.

Join points in this system are identified and defined based on OSM-based AST nodes and the control flow of the program dynamically, which is represented specific moments during the execution of EHR operations, where certain EHR actions can be performed. In this context, OSMParser and AOPVisitor join points capture the control flow of EHR operations, thereby providing necessary interaction points. In automation, Aspects within FormalProof are defined to facilitate the construction of CFG, and the subsequent translation of this graph into a formal model. The use of extended join points enables us to invoke the formal verification process at the proper stages, such as the ParseAOPSource's pointcuts intercept the 'register' and 'edit' methods associated with patients and providers in EHRs. These pointcuts implement data obliviousness and anonymization techniques. The pointcuts extend from the join point, where the processAspects() method is initiated. Furthermore, they intercept 'record' and 'retrieve' methods, providing advice and enabling the real-time integration of additional behavior and functionality without modifying the inherent structure of the program timely.

OSM's analyzer confirms that EHR system adheres to the constraints and requirements. It does this by systematically examining the outputs generated by OSM-EHR system in response to given inputs. When the system generates an output, the analyzer cross-checks this against the expectations defined by the formal model as mentioned. For instance, if EHR system states that a patient x is receiving health service support, the analyzer checks to confirm that access control and data privacy measures are also implemented for that patient. When OSM-EHR system generates output based on these inputs, such as predictions or recommendations, we cross-check it against our expectations derived from OSM's model checking. To clarify that, the logic we have derived stipulates that if a patient is receiving health service support, it is mandatory that robust data privacy and security measures are also implemented. In this context, the logical relationship, i.e., proposition 4, signifies that the presence of health service support for a patient x necessitates the implementation of both access control and data privacy measures for the same patient:

$\exists x(C(x) \Leftrightarrow A(x) \wedge B(x))$.

This relationship is confirmed using a formal mathematical model. Let us look at proposition 5:

$A(x) + B(x) \Rightarrow C(x)$.

Obviously, this means that if EHR(x) has access to MIMIC-EHR health service services, then patient x also has data privacy and security (A and B), as shown by the expression. This makes sense to us, as protecting the privacy and security of patient information during health service requires both access control and data privacy. The implication of this proposition is that if patient x has A or B, then EHR(x) must also have health service support. However, there may be exceptions to this rule, such as when access restriction or data privacy is needed for reasons other than health service assistance. A patient may have permission to see his/her EHR or data privacy concerns may prevent him/her from using the hospital system. With applying OSM's model checking, we implement the connection between C, A, and B using different operator than addition, where, $C(x) \Leftrightarrow A(x) \wedge B(x)$.

To verify this logical relationship and ensure the system's compliance with these dependencies, this means that OSM's formal mode can verify the AO privacy and security module, where if EHR(x) has health service support, then patient x must have A and B. Therefore, this guarantee that the absence of health service support does not imply the absence of access restrictions or data privacy. In this context, is an automated process that scrutinizes the system to confirm whether it correctly adheres to these defined rules. Through this process, any discrepancies or faults inside EHR system that go against the predetermined attributes, the analyzer would flag these issues. When such issues are found, they can be corrected by revising AOP code, and the model checking process can be repeated to verify the fix. To ensure that the changes made do not bring any new problems, and to verify that the system now correctly implements and interacts with its aspects, OSM's model verification is repeated. The quality of service and general dependability of the system are improved by OSM's use of a formal approach and automated verification tools, which enable early detection and resolution of possible problems.

## CONCLUSIONS

AOP-based statistical model checking generation method, using a propositional model, proves to be an invaluable tool in ensuring the safety, reliability, and adaptability of real-time embedded and safety-critical systems, especially in the continuously evolving context of EHR software systems. The propositional model aids in understanding and predicting the behavior of these systems under varying conditions, thereby providing enhanced modularity, reusability, and maintaining software integrity. It facilitates the validation of the implementation phase by effectively analyzing and processing the diverse scenarios arising from a constantly changing environment. The construction of OSM framework involves parsing the source code, creating CFGs, and translating these into formal propositional models that are then checked against specific properties to ensure that the system meets all defined requirements. This methodological approach heightens the safety, reliability, and overall quality of these systems. The application of this methodology in the context of the COVID-19 pandemic demonstrates its effectiveness in EHR system adaptations. Addressing crucial cross-cutting concerns like access control, data privacy, health service support, and vaccine management, OSM process successfully validates the system's desired properties, leading to improved reliability, adaptability, and quality of EHR system. Using formal method with AOP identified potential issues and verifying system compliance with required properties at real time. Dynamically, in cases, where discrepancies are detected, the code can be modified, and the process is repeated until OSM's system meets the desired standards. Consequently, AOP-based statistical model checking generation method ensures the correct implementation and harmonious interaction of aspects, proving its value for developers and engineers working on real-time embedded and safety-critical systems in a continuously changing environment. This enables the identification of potential issues and confirmation of system compliance with desired properties. In the case of detected flaws, the code can be modified, and the process repeated to ensure the system's correct implementation and interaction of aspects.

**Future Works & Limitations**

Future research efforts will aim to fully exploit the capabilities of AOP-based statistical model checking generation method in the context of big data. This will involve addressing issues related to scalability, applicability across various domains, data characteristics, privacy and security concerns, as well as the integration of predictive analytics. Further investigation is required to examine the scalability of this approach, as it is a crucial aspect that requires thorough examination. Considering the rapid expansion of big data, it is imperative to investigate the capacity of the proposed model to effectively accommodate and process significantly larger datasets. To achieve scalability, it is imperative to explore the potential of distributed and parallel computing techniques to enhance the computational efficiency of statistical model checking within a big data setting. It is imperative to investigate the resilience and efficacy of the model across various big data domains, extending beyond the realm of healthcare. This would enable us to comprehend the model's suitability and adaptability in various contexts, thereby augmenting its potential for wider influence. Furthermore, it is possible to further explore distinct attributes of big data, namely veracity, velocity, and variety. Subsequent research endeavors may prioritize the enhancement of real-time data processing capabilities, the assurance of data quality and accuracy, and the effective management of diverse data types, all while upholding the system's safety and reliability. In the future, it is imperative to conduct a comprehensive examination of the security and privacy considerations pertaining to the system, as further investigation is warranted. Considering the inherent sensitivity of EHRs and the increased significance of safeguarding data privacy in the current era of extensive data collection, conducting research on the deployment of sophisticated data encryption methods, secure access mechanisms, and anonymization techniques would yield valuable insights. Furthermore, considering the artificial intelligence standpoint and IoT, future investigations may prioritize the exploration of methods to augment our model by incorporating predictive analytics functionalities (Alsobeh, 2018). This approach has the potential to enhance the identification of prospective challenges and system failures, consequently enhancing the efficacy of preventive measures and augmenting the overall dependability and performance of the system. Although our study offers interesting insights into the application and effectiveness of OSM strategy, it is important to acknowledge its limits. The research was primarily limited by its narrow scope, which centered on systems of modest size. As a result, it may have failed to consider the unique issues associated with larger-scale and more complex designs. The potential computational burden associated with formal modeling may not accurately reflect the demands of real-time situations, indicating the need for care when generalizing our results to such settings.

**Funding:** The author received no financial support for the research and/or authorship of this article.

**Ethics declaration:** The author declared that, as this study proposes a conceptual model and utilizes publicly available datasets, formal ethical approval was not required. To access the dataset, a CITI course was completed in human research protections as a requirement to protect the privacy of the individuals whose data is in the database.

**Declaration of interest:** The author declares no competing interest.

**Data availability:** Data generated or analyzed during this study are available from the author on request.

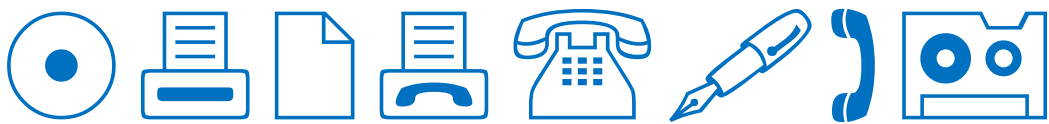